\documentclass[floatfix,secnumarabic,amssymb, nobibnotes, aps, prd]{revtex4-2}

\usepackage{subfigure}
\usepackage{graphicx}

\begin{document}


\title{A Computation-Enhanced High-Dimensional Quantum Gate for Silicon-Vacancy Spins}



\author{Gang Fan$^{1,2}$ and Fang-Fang Du$^1$}
\email[]{dufangfang19871210@163.com}

\address{$^1$Science and Technology on Electronic Test
	and Measurement Laboratory, North University of China, Taiyuan, 030051, China \\
	$^2$Institute for Quantum Science and Technology, College of Science, National University of Defense Technology, Changsha, 410073, China
}


\date{\today}

\begin{abstract}
Qudit-based quantum gates in high-dimensional Hilbert space
can provide a viable route towards effectively accelerating the speed of quantum computing
and performing complex quantum logic operations. In the paper,
we propose a  2-qudit $4\times4$-dimensional controlled-not (CNOT) gate
for four silicon-vacancy spins,
in which the first two electron-spin states in silicon-vacancy centers
are encoded as the control qudits, and the other ones as the target qudits.
The proposed protocol is implemented with assistance of an ancillary photon
that serves as a common-data bus linking four motionless silicon-vacancy spins
placed in four independent single-sided optical nanocavities.
Moreover, the CNOT gate works in a deterministic manner
by performing the relational feed-forward operations corresponding to
the diverse outcomes of the single-photon detectors to be directed against
the ancillary photon. Further, it can be potentially generalized
to other solid-state quantum system. Under current technological conditions,
both the efficiency and fidelity of the 2-qudit CNOT gate are high.
\end{abstract}


\maketitle

\section{\label{sec1}Introduction}

Quantum computing is an emerging technology that
leverages the essential principles of quantum mechanics
to execute computational tasks that are inefficient or intractable
for classical computing \cite{RN197,RN208,he2023criticality,li2024heralded,SciPostPhys021}.
Quantum logic gates \cite{SongPRA,wei-adv}
enable precise manipulation of quantum qubits
to facilitate the execution of specific computational tasks
and play a crucial part in quantum information processing (QIP) \cite{zhou2022deterministic,SciPostPhys250},
where
all unitary operations can be implemented with a set of universal
two-qubit controlled-not (CNOT) gates and parallel one-qubit operations \cite{liu2008analytic}.
Consequently, the ongoing research and advancement of the CNOT gates
emerge as pivotal catalysts propelling the evolution of quantum technology, i.e., a wide range of fascinating
applications in entanglement concentration \cite{wei-PRApplied2}, entanglement purification \cite{Sheng-PhysRevA,yan2022measurement}, 
quantum secure direct communication \cite{ye2021generic,Sheng-PRApplied}, and quantum key distribution \cite{yan2021measurement}.

The traditional CNOT gates only perform logic operations on two-dimensional qubits, i.e., $|0\rangle$ and $|1\rangle$,
and
the relational transformations are
$|0_{c}, 0_{t}$$\rangle\rightarrow$$|0_{c}, 0_{t}\rangle$,
$|0_{c}, 1_{t}$$\rangle\rightarrow|0_{c}, 1_{t}\rangle$,
$|1_{c}, 0_{t}$$\rangle\rightarrow$$|1_{c}, 1_{t}\rangle$, and
$|1_{c}, 1_{t}$$\rangle\rightarrow$$|1_{c}, 0_{t}\rangle$.
Based on the two-dimensional qubit system,
a variety of quantum algorithms may be executed with the help of special quantum circuits
composed of basic gates corresponding to
unitary matrices and further require  extensive qubits
to encode information, which limit the flexibility of quantum computing
to face the challenge of large-scale integration in practice.
To overcome the limitation, high-dimensional quantum logic gates \cite{He16,AA,PhysRevLettGates,Xgate,WangTJ2,Wei-npj,Wei-PRApplied}
that exploit high-dimensional quantum qudits to encode and process multi-valued information,
have been proposed recently, thus
improving the speed of quantum computing and QIP.
Besides, high-dimensional systems not only are flexible with regard to improvements to the channel capacity and security requiring less resource overheads, but also improve the execution of quantum algorithms to offer enhanced computational capacity and higher computational accuracy.


Extensively and in-depth studied have been devoted to have been devoted to  qubit-based CNOT gates in both theory and experiment with diverse physical systems, such as linear optics \cite{photon1,photon2,wei-PRApplied3}, superconducting qubits \cite{superconducting1,superconducting2}, nuclear magnetic resonance \cite{NMR1,NMR2}, atoms \cite{Duan,Isenhower,Reisterer2,photon-gate}, color centers in diamond \cite{WangTJ3,weiNV1,weiNV2}, and quantum dots \cite{spin1,spin2,weihrspin3,SciPostPhys089}.
Among these systems, the silicon-vacancy
(SiV) center has been regarded as one of
the attractive candidates to implement scalable general-purpose QIP due to excellent optical
property and scalability with long coherence time.
Especially,
the SiV center appears a like four-level system \cite{PRA042313,PRL263601,RN204,PRL063601,Lemonde_2019},
making it suitable for variable interesting tasks, i.e.,  entanglement generation of multiple SiV centers
and the realization of tunable and strong coupling between photons and SiV centers located in the single quantum
level.
Establishing a connection between an auxiliary photon and individual spins
is a fundamental requirement for the development of QIP \cite{NV1}.
A deterministic link can be achieved by employing nanocavity-SiV$^{-}$ system.

In the paper,
we investigate the implementation of the deterministic  2-qudit $4\times4$-dimensional CNOT gate for a four-
SiV system,
in which both the control and target qudits are encoded on two SiV spins.
An ancillary photon is available for contacting and interacting with four stationary SiV spins
embedded independently in four single-sided optical cavities. After that,
the photon is detected by the single-photon detectors and then the SiV spins are
executed the relational feed-forward operations to make the CNOT gate with unity success
probability in principle.

\section{\label{sec2}An silicon-vacancy center}

\begin{figure}[htbp]
	\centering
	\includegraphics[width=0.5\linewidth]{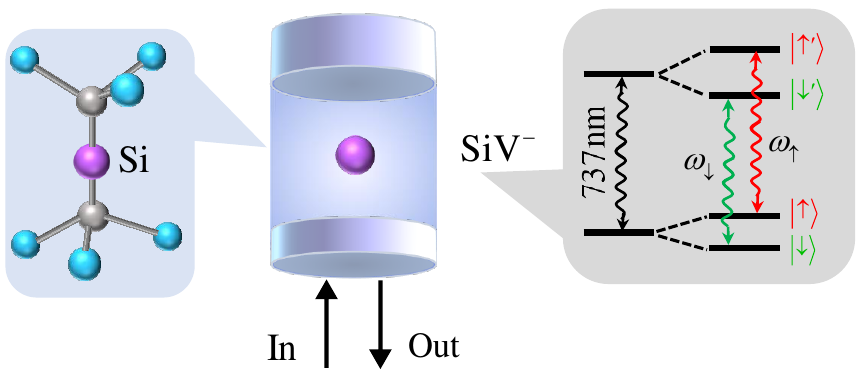}
	\caption{(a) The coupling system involved a SiV$^{-}$
		center and an optical cavity; (b) The structural arrangement of energy levels and optical transitions within the SiV$^{-}$ center.	
	} \label{fig1}
\end{figure}

A negatively charged SiV (SiV$^{-}$) center
configuration placed in a single-sided nanocavity with a
resonance frequency to be close to 737 nm as depicted in Figure \ref{fig1}.
A substitutional silicon atom is located between two carbon vacancies of the SiV$^{-}$ center  \cite{RN199,PRB165428,RN200}.
The consequential $D_{3d}$ inversion
symmetry of the SiV$^{-}$ center with regard to the Si atom  brings about a
disappearing electron-dipole moments of the excited and ground
states \cite{RN199}, so the SiV$^{-}$ center is noise-resistant, especially
fixed in nanophotonic framework,
and has better optical performances than a nitrogen-vacancy (NV)
center facing a low temperature (lower than 500 mK) \cite{PRB165428}. Under moderate strain, the energy
levels of the SiV$^{-}$ center
are reduced to two spin sublevels in the ground (exited) states, i.e., $|\downarrow\rangle$  and $|\uparrow\rangle$
($|\downarrow'\rangle$ and $|\uparrow'\rangle$),
ignoring other unrelated levels due to large detunings.
If an external effective magnetic
field is introduced along symmetry axis of the SiV$^{-}$ center, two spin-conserving
transitions, i.e., $|\downarrow\rangle\rightarrow |\downarrow'\rangle$ and
$|\uparrow\rangle\rightarrow |\uparrow'\rangle$
with respective resonant frequencies $\omega_{\downarrow}$ and $\omega_{\uparrow}=\omega_{\downarrow}+\Delta$
can link these
states with horizontal polarization, namely, $|H\rangle$,
but
the two cross transitions corresponding to spin-flipping states are dipole forbidden
and thus they are neglected \cite{PRB165428}.
Here, $\Delta$ represents the energy difference
between the states $|\downarrow\rangle$ and $|\uparrow\rangle$ caused by the applied magnetic field.

Assuming that an photon in polarized state $|H\rangle$ with frequency $\omega$
incorporates into
the SiV$^{-}$-nanocavity system, where its photonic frequency $\omega$
is near resonance with the nanocavity one $\omega_{c}$, i.e., $\omega\simeq\omega_{c}$, resulting in
the dipole-allowed transition $|\downarrow\rangle\rightarrow |\downarrow'\rangle$
due to destructive interference,
but extremely detuned from the other
one $|\uparrow\rangle\rightarrow |\uparrow'\rangle$.
As a result, when the SiV$^{-}$ center is initialized to various ground states $|\downarrow\rangle$ and $|\uparrow\rangle$
and is coupled strongly to the nanocavity,
the dipole-allowed scattering are
different.
Moreover, the dynamic equations of
motion for the dipole lower operator $ \hat{\sigma}_{-} $ of the SiV$^{-}$ center and the
annihilation operator $\hat{a}$ of the cavity filed
in conjunction with the standard input-output relation \cite{PRB085307,RN212} can be expressed as
\begin{eqnarray} \label{eq1}
	\frac{d\hat{a}}{dt}&=&-[i(\omega_{c}-\omega)+\frac{\kappa}{2}]\hat{a}-g\hat{\sigma}_{-}-\sqrt{\kappa}\hat{a}_{in}+\hat{N},\nonumber\\
	\frac{d\hat{\sigma}_{-}}{dt}&=&-[i(\omega_{d}-\omega)+\frac{\gamma}{2}]\hat{\sigma}_{-}-g\hat{\sigma}_{z}\hat{a}+\hat{N^{\prime}},\nonumber\\
	\hat{a}_{out}&=&\hat{a}_{in}+\sqrt{\kappa}\hat{a}.
\end{eqnarray}
where $\omega_{d}(d=\downarrow,\uparrow)$ is the dipole-allowed transition frequency, and
$\kappa$ and $\gamma$ signify the decay rates of the cavity field and the excited states of the SiV$^{-}$ center.
$g$ represents the coupling ratio between the nanocavity filed and the SiV$^{-}$ center.
The noise operators
$\hat{N}$ and $\hat{N^{\prime}}$ due to the cavity leakage and dipole decay,
respectively, are introduced to keep the wanted commutation relations.
$\hat{a}_{in}$ and $\hat{a}_{out}$ represent the operators associated with the input and output vacuum fields, respectively.
Additionally, $\hat{\sigma}_{z}$
serves as the population operator for the input photon.

With weak excitation limitation ($\langle\hat{\sigma}_{z}\rangle=-1$),
the reflection coefficient, labeled as $r_{s}(\omega)$, when an photon enter and interact with the one-sided nanocavity-SiV$^{-}$ system can be accurately expressed as \cite{litao}
\begin{eqnarray}\label{eq2}
	r_{d}(\omega)=1-\frac{2(1+ i\Delta_{d})}{C+(1+ i\Delta_{d})(1+ i\Delta_{c})},
\end{eqnarray}
the subscript $d=\uparrow (\downarrow)$ of the reflection coefficient $r_{d}(\omega)$
indicates the interaction of the polarized photon with the cavity-SiV$^{-}$ system.
$C=4g^{2}/\kappa\gamma$ represents the cooperativity.  $\Delta_{d}=2(\omega_{d}-\omega)/\gamma$ and $\Delta_{c}=2(\omega_{c}-\omega)/\kappa$
denote the effective detunings of the dipole-transition and cavity-mode frequencies to be aimed at the input-field frequency $\omega$.
Consequently,
the input-output relations of the $H$-polarized photon interacting with one-sided nanocavity-SiV$^{-}$-center system can be
obtained as
$|H\rangle|\downarrow\rangle \rightarrow r_{\downarrow}(\omega)|H\rangle|\downarrow\rangle$ and
$|H\rangle|\uparrow\rangle \rightarrow r_{\uparrow}(\omega)|H\rangle|\uparrow\rangle$.
For $\Delta_{c}=\Delta_{\downarrow}=0$ and $\Delta_{\uparrow}\gg C \gg 1$,
$r_{\downarrow}=1$ and $r_{\uparrow}=-1$ for
the SiV$^{-}$ center in the various ground states $|\downarrow\rangle$
and $|\uparrow\rangle$ for the $H$-polarized photon
can be obtained.
In essence, the reflection coefficient $r_{d}(\omega)$ is governed by the spin state of the SiV$^{-}$ center \cite{PRA022428}.

\section{A  2-qudit $4\times4$-dimensional CNOT gate} \label{sec3}

\begin{figure}
	\centering
	\includegraphics[width=0.98\linewidth]{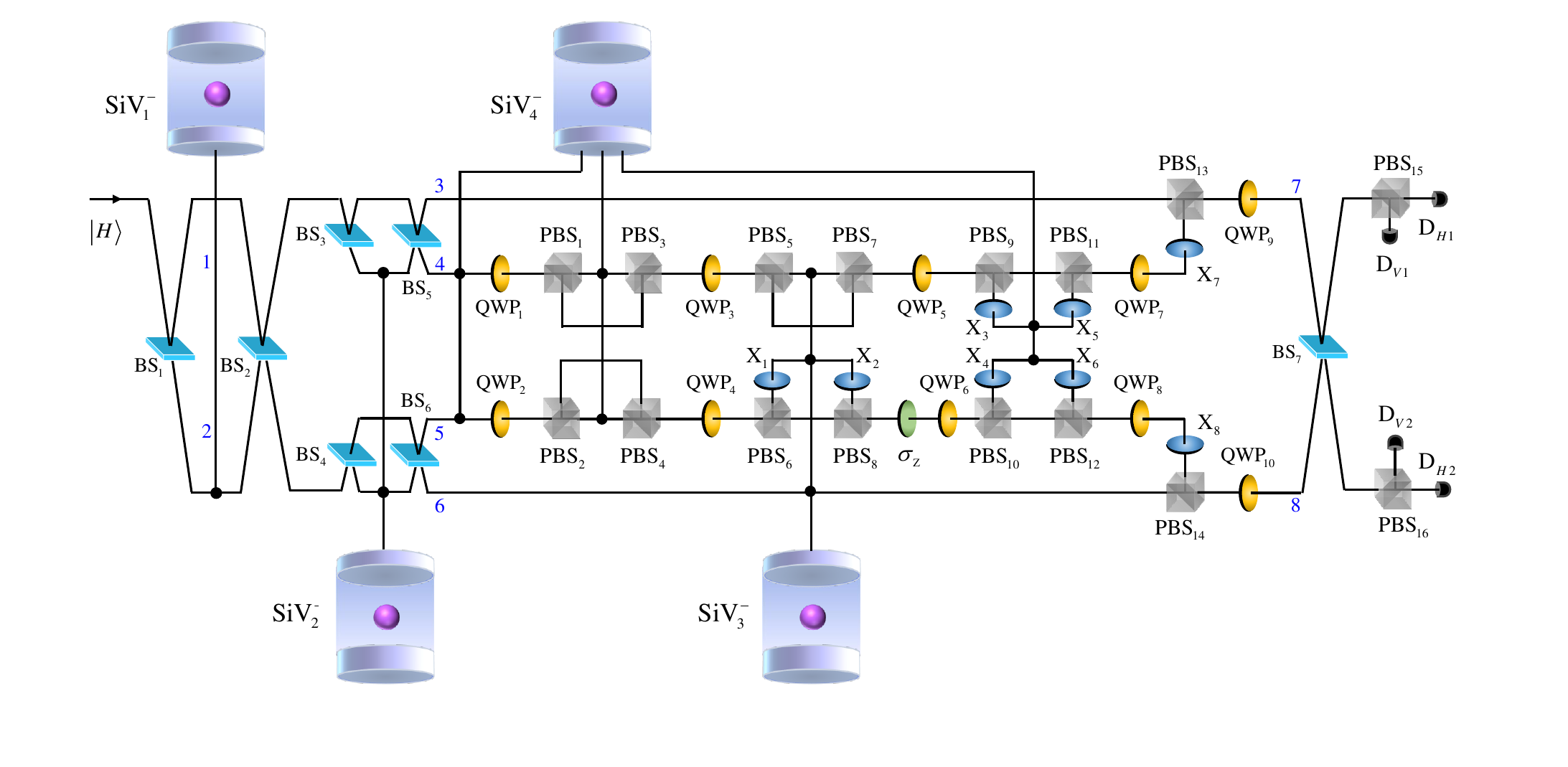}
	\caption{\label{fig2} Schematic diagram of the  2-qudit $4\times4$-dimensional CNOT gate for SiV$^{-}$ centers.
		The beam splitter (BS$_{n}, n=1,2,...,7)$ completes  path transformation between down $(d)$ and up $(u)$ ones, i.e.,
		$N_{u}\rightarrow(N_{u}+N_{d})/\sqrt{2}$ and $N_{d}\rightarrow(N_{u}-N_{d})/\sqrt{2}, N\in \{H,V\}$.
		The polarized beam splitter (PBS$_{i}, i=1,2,...,16$) transmits (reflects)  $|H\rangle$-($|V\rangle$)-polarized component of one photon.
		The X$_{j}(j=1,2,...,4)$ fulfills the bit-flip operation $|H\rangle\leftrightarrow|V\rangle$.
		The quarter-wave plate (QWP$_{k}, k=1,2...,10)$ fulfills the Hadamard operation. $\rm \sigma_{z}$ completes the phase-flip operation $\sigma_{z}=-|H\rangle\langle H|-|V\rangle\langle V|$. The SiV$^{-}_{m}( m=1,2,...,4)$ represents the SiV spin.
		The single photon detector ($\rm D_{H1}$, $\rm D_{V1}$, $\rm D_{H2}$, or $\rm D_{V2}$) can be used to detect the auxiliary photon.
	}
\end{figure}

Now we set up the deterministic  2-qudit 4$\times$4-dimensional CNOT gate for a four-SiV$^{-}$ system
with a auxiliary photon in Figure \ref{fig2}.
Here, we encode spin states
$|\downarrow\downarrow\rangle_{12}$, $|\downarrow\uparrow\rangle_{12}$,
$|\uparrow\downarrow\rangle_{12}$, and $|\uparrow\uparrow\rangle_{12}$
of two SiV$^{-}_{1}$ and SiV$^{-}_{2}$ centers
as control qudits
$|0\rangle_{c}$, $|1\rangle_{c}$, $|2\rangle_{c}$, and $|3\rangle_{c}$, respectively, and the other ones
$|\downarrow\downarrow\rangle_{34}$, $|\downarrow\uparrow\rangle_{34}$,
$|\uparrow\downarrow\rangle_{34}$, and $|\uparrow\uparrow\rangle_{34}$ of SiV$^{-}_{3}$ and SiV$^{-}_{4}$ centers
as the target qudits $|0\rangle_{t}$, $|1\rangle_{t}$, $|2\rangle_{t}$, and $|3\rangle_{t}$, respectively.
The  2-qudit $4\times4$-dimensional CNOT gate
makes the target qudit in state $|t\rangle$ to become $|(c+t)\%d\rangle$
when the control qudit is in state $|c\rangle$, as the following transformation \cite{PhysRevLettGates}:
\begin{eqnarray} \label{eq3}
	|0_{c}, 0_{t}\rangle\rightarrow|0_{c}, 0_{t}\rangle,\quad  |0_{c}, 1_{t}\rangle\rightarrow|0_{c}, 1_{t}\rangle, 
	|0_{c}, 2_{t}\rangle\rightarrow|0_{c}, 2_{t}\rangle, \quad  |0_{c}, 3_{t}\rangle\rightarrow|0_{c}, 3_{t}\rangle,   \nonumber \\
	|1_{c}, 0_{t}\rangle\rightarrow|1_{c}, 1_{t}\rangle, \quad |1_{c}, 1_{t}\rangle\rightarrow|1_{c}, 2_{t}\rangle, 
	|1_{c}, 2_{t}\rangle\rightarrow|1_{c}, 3_{t}\rangle, \quad |1_{c}, 3_{t}\rangle\rightarrow|1_{c}, 0_{t}\rangle,   \nonumber \\
	|2_{c}, 0_{t}\rangle\rightarrow|2_{c}, 2_{t}\rangle, \quad  |2_{c}, 1_{t}\rangle\rightarrow|2_{c}, 3_{t}\rangle,
	|2_{c}, 2_{t}\rangle\rightarrow|2_{c}, 0_{t}\rangle, \quad  |2_{c}, 3_{t}\rangle\rightarrow|2_{c}, 1_{t}\rangle,   \nonumber \\
	|3_{c}, 0_{t}\rangle\rightarrow|3_{c}, 3_{t}\rangle, \quad  |3_{c}, 1_{t}\rangle\rightarrow|3_{c}, 0_{t}\rangle, 
	|3_{c}, 2_{t}\rangle\rightarrow|3_{c}, 1_{t}\rangle, \quad  |3_{c}, 3_{t}\rangle\rightarrow|3_{c}, 2_{t}\rangle.   
\end{eqnarray}

Suppose that the control two spins of two SiV$^{-}_{1}$ and SiV$^{-}_{2}$ centers
is originally prepared in state
$|\psi\rangle_{c}=\alpha_{1}|3\rangle_{c}+\alpha_{2}|2\rangle_{c}+\alpha_{3}|1\rangle_{c}+\alpha_{4}|0\rangle_{c}$  and the control two spins of two SiV$^{-}_{3}$ and SiV$^{-}_{4}$ centers
is originally prepared in state $|\psi\rangle_{t}=\gamma_{1}|3\rangle_{t}+\gamma_{2}|2\rangle_{t}+\gamma_{3}|1\rangle_{t}+\gamma_{4}|0\rangle_{t}$, respectively.
Here, $|B_{1}|^{2}+|B_{2}|^{2}+|B_{3}|^{2}+|B_{4}|^{2}=1$, $(B=\alpha, \gamma)$.
The ancillary single photon is originally prepared in state $|H\rangle$.
Thus, the state of the whole system  involving the single photon and four SiV$^{-}$ centers is $|\Psi\rangle_{0}=|H\rangle\otimes|\psi\rangle_{c}\otimes|\psi\rangle_{t}$.

Firstly, the single photon experiences some operations as follows: $\rm BS_{1} \rightarrow \rm SiV_{1}^{-}$-cavity system $\rightarrow\rm BS_{2}\rightarrow \rm BS_{3}, \rm BS_{4}\rightarrow\rm SiV_{2}^{-}$-cavity system $\rightarrow \rm BS_{5}, BS_{6}$. Here,
the optical element beam splitter (BS$_{n}, n=1,2,...,6)$ completes path transformations between down $(d)$ and up $(u)$ ones, i.e.,
$H_{u}\rightarrow(H_{u}+H_{d})/\sqrt{2}$ and $H_{d}\rightarrow(H_{u}-H_{d})/\sqrt{2}$.
After these operations, the state $|\Psi\rangle_{0}$ of the whole system is evolved into
\begin{eqnarray} \label{eq4}
	|\Psi\rangle_{1}&=& (\alpha_{1}|H\rangle^{5}|3\rangle_{c}+\alpha_{2}|H\rangle^{6}|2\rangle_{c}+\alpha_{3}|H\rangle^{4}|1\rangle_{c}+\alpha_{4}|H\rangle^{3}|0\rangle_{c})
	\nonumber \\
	&& \otimes(\gamma_{1}|3\rangle+\gamma_{2}|2\rangle+\gamma_{3}|1\rangle+\gamma_{4}|0\rangle)_{t}.
\end{eqnarray}

Secondly, before and after the auxiliary photon in path 4 (path 5) interacts with $\rm SiV_{4}^{-}$-cavity system, the
Hadamard operation $\rm H_{e}$ aiming at the spin of the $\rm SiV_{4}^{-}$ by taking advantage of a $\pi/2$ microwave pulse completes the transformations $|+\rangle\rightarrow\frac{1}{\sqrt{2}}(|+\rangle+|-\rangle)$ and $|-\rangle\rightarrow\frac{1}{\sqrt{2}}(|+\rangle-|-\rangle)$.
After the first interaction between the auxiliary photon and the $\rm SiV_{4}^{-}$-cavity system,
the state $|\Psi\rangle_{1}$ of the whole system is changed into
\begin{eqnarray} \label{eq5}
	|\Psi\rangle_{2}&=&
	\alpha_{1}|H\rangle^{5}|3\rangle_{c} \otimes (\gamma_{1}|2\rangle+\gamma_{2}|3\rangle+\gamma_{3}|0\rangle+\gamma_{4}|1\rangle)_{t}   \nonumber \\
	&&\!+\alpha_{2}|H\rangle^{6}|2\rangle_{c} \otimes (\gamma_{1}|3\rangle+\gamma_{2}|2\rangle+\gamma_{3}|1\rangle+\gamma_{4}|0\rangle)_{t}  \nonumber \\
	&&\!+\alpha_{3}|H\rangle^{4}|1\rangle_{c}	\otimes (\gamma_{1}|2\rangle+\gamma_{2}|3\rangle+\gamma_{3}|0\rangle+\gamma_{4}|1\rangle)_{t}  \nonumber \\
	&&\!+\alpha_{4}|H\rangle^{3}|0\rangle_{c} \otimes (\gamma_{1}|3\rangle+\gamma_{2}|2\rangle+\gamma_{3}|1\rangle+\gamma_{4}|0\rangle)_{t}.
\end{eqnarray}

Thirdly, the photon on path 4 (path 5) experiences following operations: $\rm QWP_{1} (\rm QWP_{2})\rightarrow$ $\rm PBS_{1} (\rm PBS_{2})$ $\rightarrow$ $\rm SiV_{4}^{-}$-cavity system $\rightarrow$ $\rm PBS_{3}$ ($ \rm PBS_{4})$$\rightarrow$ $\rm QWP_{3}$ ($ \rm QWP_{4})$, where the polarized beam splitter (PBS$_{i}, i=1,2,...,4$) transmits (reflects)  $|H\rangle$-($|V\rangle$)-polarized component of the photon and meanwhile the optical element quarter-wave plate $\rm QWP_{k} (k=1,2...,4)$ is to perform the Hadamard operation on the auxiliary photon, i.e., $|H\rangle\rightarrow\frac{1}{\sqrt{2}}(|H\rangle+|V\rangle)$ and $|V\rangle\rightarrow\frac{1}{\sqrt{2}}(|H\rangle-|V\rangle)$.
After the second interaction between the auxiliary photon and the $\rm SiV_{4}^{-}$-cavity system,
the state $|\Psi\rangle_{2}$ of the whole system is changed into
\begin{eqnarray} \label{eq6}
	|\Psi\rangle_{3}&=&
	\alpha_{1}|H\rangle^{5}|3\rangle_{c} \otimes (\gamma_{1}|2\rangle+\gamma_{3}|0\rangle)_{t}   -\alpha_{1}|V\rangle^{5}|3\rangle_{c} \otimes (\gamma_{2}|3\rangle+\gamma_{4}|1\rangle)_{t}   \nonumber \\
	&&\!+\alpha_{2}|H\rangle^{6}|2\rangle_{c} \otimes (\gamma_{1}|3\rangle+\gamma_{2}|2\rangle+\gamma_{3}|1\rangle+\gamma_{4}|0\rangle)_{t}  \nonumber \\
	&&\!+\alpha_{3}|H\rangle^{4}|1\rangle_{c}	\otimes (\gamma_{1}|2\rangle+\gamma_{3}|0\rangle)_{t} -\alpha_{3}|V\rangle^{4}|1\rangle_{c}	\otimes (\gamma_{2}|3\rangle+\gamma_{4}|1\rangle)_{t}   \nonumber \\
	&&\!+\alpha_{4}|H\rangle^{3}|0\rangle_{c} \otimes (\gamma_{1}|3\rangle+\gamma_{2}|2\rangle+\gamma_{3}|1\rangle+\gamma_{4}|0\rangle)_{t}.
\end{eqnarray}

Fourthly, the photon on path 4 (path 5) experiences following operations: $\rm PBS_{5}$ ($ \rm PBS_{6}, X_{1})\rightarrow$ $\rm SiV_{3}^{-}$-cavity system $\rightarrow$ $\rm PBS_{7} $ ($\rm X_{2}, PBS_{8}, \sigma_{z})$, meanwhile, the photon on path 6 also interacts with $\rm SiV_{3}^{-}$-cavity system. Here, the optical element X$_{j}(j=1,2,...,4)$ is a half-wave plate fixed at $45^{\circ}$ to
fulfill a bit-flip operation  $|H\rangle\leftrightarrow|V\rangle$ and $\rm \sigma_{z}$ completes the phase-flip operation $\sigma_{z}=-|H\rangle\langle H|-|V\rangle\langle V|$.
Notably, before and after the photon experiences these operations,
$H^{e}$ should be acted on the spin of $\rm SiV_{3}^{-}$ center,
resulting in
\begin{eqnarray} \label{eq7}
	|\Psi\rangle_{4}&=&
	\alpha_{1}|H\rangle^{5}|3\rangle_{c} \otimes (\gamma_{1}|2\rangle+\gamma_{3}|0\rangle)_{t}   +\alpha_{1}|V\rangle^{5}|3\rangle_{c} \otimes (\gamma_{2}|1\rangle+\gamma_{4}|3\rangle)_{t}   \nonumber \\
	&&\!+\alpha_{2}|H\rangle^{6}|2\rangle_{c} \otimes (\gamma_{1}|1\rangle+\gamma_{2}|0\rangle+\gamma_{3}|3\rangle+\gamma_{4}|2\rangle)_{t}  \nonumber \\
	&&\!+\alpha_{3}|H\rangle^{4}|1\rangle_{c}	\otimes 
	(\gamma_{1}|0\rangle+\gamma_{3}|2\rangle)_{t}-\alpha_{3}|V\rangle^{4}|1\rangle_{c}	\otimes (\gamma_{2}|3\rangle+\gamma_{4}|1\rangle)_{t}   \nonumber \\
	&&\!+\alpha_{4}|H\rangle^{3}|0\rangle_{c} \otimes (\gamma_{1}|3\rangle+\gamma_{2}|2\rangle+\gamma_{3}|1\rangle+\gamma_{4}|0\rangle)_{t}.
\end{eqnarray}

Fifthly, the photon on path 4 (path 5) experiences following operations: $\rm QWP_{5}$ ($\rm QWP_{6})$ $\rightarrow$ $\rm PBS_{9} $ ($\rm PBS_{10}$) $\rightarrow \rm X_{3} $ ($\rm X_{4}$) $\rightarrow \rm SiV_{4}^{-}$-cavity system $\rightarrow \rm X_{5} $ ($\rm X_{6}$) $\rightarrow$ $\rm PBS_{11}$ ($\rm PBS_{12}) \rightarrow$ $\rm QWP_{7}$ ($\rm QWP_{8})
\rightarrow$ $\rm X_{7}$ ($\rm X_{8}$)$\rightarrow$ $\rm PBS_{13}$ ($\rm PBS_{14})$ mixed at path 7 (8).
After the third interaction between the auxiliary photon and the $\rm SiV_{4}^{-}$-cavity system,
leading to
\begin{eqnarray} \label{eq8}
	|\Psi\rangle_{5}&=&
	\alpha_{1}|V\rangle^{8}|3\rangle_{c} \otimes (\gamma_{1}|2\rangle+\gamma_{2}|1\rangle+\gamma_{3}|0\rangle+\gamma_{4}|3\rangle)_{t}   \nonumber \\
	&&\!+\alpha_{2}|H\rangle^{8}|2\rangle_{c} \otimes (\gamma_{1}|1\rangle+\gamma_{2}|0\rangle+\gamma_{3}|3\rangle+\gamma_{4}|2\rangle)_{t}  \nonumber \\
	&&\!+\alpha_{3}|V\rangle^{7}|1\rangle_{c}	\otimes (\gamma_{1}|0\rangle+\gamma_{2}|3\rangle+\gamma_{3}|2\rangle+\gamma_{4}|1\rangle)_{t}  \nonumber \\
	&&\!+\alpha_{4}|H\rangle^{7}|0\rangle_{c} \otimes (\gamma_{1}|3\rangle+\gamma_{2}|2\rangle+\gamma_{3}|1\rangle+\gamma_{4}|0\rangle)_{t}.
\end{eqnarray}

Finally, the photon on path 7 (path 8) experiences following operations: $\rm QWP_{9}$  ($\rm QWP_{10}$)$\rightarrow$ $\rm BS_{7}\rightarrow$ $\rm PBS_{15}$ ($\rm PBS_{16})$, resulting in
\begin{eqnarray} \label{eq9}
	|\Psi\rangle_{6}&\!=\!&
	|\rm D_{H1}\rangle\otimes
	[\alpha_{1}|3\rangle_{c} \otimes (\gamma_{1}|2\rangle+\gamma_{2}|1\rangle+\gamma_{3}|0\rangle+\gamma_{4}|3\rangle)_{t}   \nonumber \\
	&&+\alpha_{2}|2\rangle_{c} \otimes (\gamma_{1}|1\rangle+\gamma_{2}|0\rangle+\gamma_{3}|3\rangle+\gamma_{4}|2\rangle)_{t}  \nonumber \\
	&&+\alpha_{3}|1\rangle_{c}	\otimes (\gamma_{1}|0\rangle+\gamma_{2}|3\rangle+\gamma_{3}|2\rangle+\gamma_{4}|1\rangle)_{t}  \nonumber \\
	&&+\alpha_{4}|0\rangle_{c} \otimes (\gamma_{1}|3\rangle+\gamma_{2}|2\rangle+\gamma_{3}|1\rangle+\gamma_{4}|0\rangle)_{t}]  \nonumber \\
	&&+|\rm D_{V1}\rangle\otimes
	[-\alpha_{1}|3\rangle_{c} \otimes (\gamma_{1}|2\rangle+\gamma_{2}|1\rangle+\gamma_{3}|0\rangle+\gamma_{4}|3\rangle)_{t}   \nonumber \\
	&&-\alpha_{2}|2\rangle_{c} \otimes (\gamma_{1}|1\rangle+\gamma_{2}|0\rangle+\gamma_{3}|3\rangle+\gamma_{4}|2\rangle)_{t}  \nonumber \\
	&&+\alpha_{3}|1\rangle_{c}	\otimes (\gamma_{1}|0\rangle+\gamma_{2}|3\rangle+\gamma_{3}|2\rangle+\gamma_{4}|1\rangle)_{t}  \nonumber \\
	&&+\alpha_{4}|0\rangle_{c} \otimes (\gamma_{1}|3\rangle+\gamma_{2}|2\rangle+\gamma_{3}|1\rangle+\gamma_{4}|0\rangle)_{t}]  \nonumber \\
	&&+|\rm D_{H2}\rangle\otimes
	[-\alpha_{1}|3\rangle_{c} \otimes (\gamma_{1}|2\rangle+\gamma_{2}|1\rangle+\gamma_{3}|0\rangle+\gamma_{4}|3\rangle)_{t}   \nonumber \\
	&&+\alpha_{2}|2\rangle_{c} \otimes (\gamma_{1}|1\rangle+\gamma_{2}|0\rangle+\gamma_{3}|3\rangle+\gamma_{4}|2\rangle)_{t}  \nonumber \\
	&&-\alpha_{3}|1\rangle_{c}	\otimes (\gamma_{1}|0\rangle+\gamma_{2}|3\rangle+\gamma_{3}|2\rangle+\gamma_{4}|1\rangle)_{t}  \nonumber \\
	&&+\alpha_{4}|0\rangle_{c} \otimes (\gamma_{1}|3\rangle+\gamma_{2}|2\rangle+\gamma_{3}|1\rangle+\gamma_{4}|0\rangle)_{t}]  \nonumber \\
	&&+|\rm D_{V2}\rangle\otimes
	[\alpha_{1}|3\rangle_{c} \otimes (\gamma_{1}|2\rangle+\gamma_{2}|1\rangle+\gamma_{3}|0\rangle+\gamma_{4}|3\rangle)_{t}   \nonumber \\
	&&-\alpha_{2}|2\rangle_{c} \otimes (\gamma_{1}|1\rangle+\gamma_{2}|0\rangle+\gamma_{3}|3\rangle+\gamma_{4}|2\rangle)_{t}  \nonumber \\
	&&-\alpha_{3}|1\rangle_{c}	\otimes (\gamma_{1}|0\rangle+\gamma_{2}|3\rangle+\gamma_{3}|2\rangle+\gamma_{4}|1\rangle)_{t}  \nonumber \\
	&&+\alpha_{4}|0\rangle_{c} \otimes (\gamma_{1}|3\rangle+\gamma_{2}|2\rangle+\gamma_{3}|1\rangle+\gamma_{4}|0\rangle)_{t}].
\end{eqnarray}

In evidence, if the detector $\rm D_{H1}$ responses, the standard  2-qudit $4\times4$-dimensional CNOT gate acting on four $\rm SiV^{-}$ centers can be established successfully, i.e.,
\begin{eqnarray} \label{eq10}
	|\Phi_{0}\rangle&=&
	\alpha_{1}|3\rangle_{c} \otimes (\gamma_{1}|2\rangle+\gamma_{2}|1\rangle+\gamma_{3}|0\rangle+\gamma_{4}|3\rangle)_{t}   \nonumber \\
	&&+\alpha_{2}|2\rangle_{c} \otimes (\gamma_{1}|1\rangle+\gamma_{2}|0\rangle+\gamma_{3}|3\rangle+\gamma_{4}|2\rangle)_{t}  \nonumber \\
	&&+\alpha_{3}|1\rangle_{c}	\otimes (\gamma_{1}|0\rangle+\gamma_{2}|3\rangle+\gamma_{3}|2\rangle+\gamma_{4}|1\rangle)_{t}  \nonumber \\
	&&+\alpha_{4}|0\rangle_{c} \otimes (\gamma_{1}|3\rangle+\gamma_{2}|2\rangle+\gamma_{3}|1\rangle+\gamma_{4}|0\rangle)_{t}.
\end{eqnarray}
Otherwise,
if the detectors $\rm D_{V1}$ (or $\rm D_{H2}$, $\rm D_{V2}$) responses, the nonstandard 2-qudit CNOT gate can be obtained, i.e.,
\begin{eqnarray} \label{eq11}
	|\Phi_{1}\rangle&=&
	-\alpha_{1}|3\rangle_{c} \otimes (\gamma_{1}|2\rangle+\gamma_{2}|1\rangle+\gamma_{3}|0\rangle+\gamma_{4}|3\rangle)_{t}   \nonumber \\
	&&-\alpha_{2}|2\rangle_{c} \otimes (\gamma_{1}|1\rangle+\gamma_{2}|0\rangle+\gamma_{3}|3\rangle+\gamma_{4}|2\rangle)_{t}  \nonumber \\
	&&+\alpha_{3}|1\rangle_{c}	\otimes (\gamma_{1}|0\rangle+\gamma_{2}|3\rangle+\gamma_{3}|2\rangle+\gamma_{4}|1\rangle)_{t}  \nonumber \\
	&&+\alpha_{4}|0\rangle_{c} \otimes (\gamma_{1}|3\rangle+\gamma_{2}|2\rangle+\gamma_{3}|1\rangle+\gamma_{4}|0\rangle)_{t}, \nonumber \\
	|\Phi_{2}\rangle&=&
	-\alpha_{1}|3\rangle_{c} \otimes (\gamma_{1}|2\rangle+\gamma_{2}|1\rangle+\gamma_{3}|0\rangle+\gamma_{4}|3\rangle)_{t}   \nonumber \\
	&&+\alpha_{2}|2\rangle_{c} \otimes (\gamma_{1}|1\rangle+\gamma_{2}|0\rangle+\gamma_{3}|3\rangle+\gamma_{4}|2\rangle)_{t}  \nonumber \\
	&&-\alpha_{3}|1\rangle_{c}	\otimes (\gamma_{1}|0\rangle+\gamma_{2}|3\rangle+\gamma_{3}|2\rangle+\gamma_{4}|1\rangle)_{t}  \nonumber \\
	&&+\alpha_{4}|0\rangle_{c} \otimes (\gamma_{1}|3\rangle+\gamma_{2}|2\rangle+\gamma_{3}|1\rangle+\gamma_{4}|0\rangle)_{t}, \nonumber \\
	|\Phi_{3}\rangle&=&
	\alpha_{1}|3\rangle_{c} \otimes (\gamma_{1}|2\rangle+\gamma_{2}|1\rangle+\gamma_{3}|0\rangle+\gamma_{4}|3\rangle)_{t}   \nonumber \\
	&&-\alpha_{2}|2\rangle_{c} \otimes (\gamma_{1}|1\rangle+\gamma_{2}|0\rangle+\gamma_{3}|3\rangle+\gamma_{4}|2\rangle)_{t}  \nonumber \\
	&&-\alpha_{3}|1\rangle_{c}	\otimes (\gamma_{1}|0\rangle+\gamma_{2}|3\rangle+\gamma_{3}|2\rangle+\gamma_{4}|1\rangle)_{t}  \nonumber \\
	&&+\alpha_{4}|0\rangle_{c} \otimes (\gamma_{1}|3\rangle+\gamma_{2}|2\rangle+\gamma_{3}|1\rangle+\gamma_{4}|0\rangle)_{t}.
\end{eqnarray}
Apparently, the corresponding single-qubit feed-forward operation $\sigma_{z}=|\downarrow\rangle\langle\downarrow|-|\uparrow\rangle\langle\uparrow|$
with a $\pi$ microwave pulse
is performed on $\rm SiV_{1}^{-}$ (or $\rm SiV_{2}^{-}$) center, the state $|\Phi_{1}\rangle$ (or $|\Phi_{2}\rangle$ ) can be changed into the one in Eq. (\ref{eq10}).
In detail,
for the case that $\rm D_{V1}$ is triggered, a $\sigma_{z}$ is acted on the $\rm SiV_{2}^{-}$ center;
for the case that $\rm D_{H2}$ is triggered, a $-\sigma_{z}$ is acted on the $\rm SiV_{1}^{-}$ center.
Besides,
for the case that $\rm D_{V2}$ is triggered, another  auxiliary $H$-polarization photon interact with the $\rm SiV_{1}^{-}$ and $\rm SiV_{2}^{-}$ again, we can change the state $|\Phi_{3}\rangle$ to the one $|\Phi_{0}\rangle$ in Eq. (\ref{eq10}).
As a consequence, we get the  deterministic 2-qudit CNOT gate with unity success probability in
principle on account of the reflection-selective rule of the ancillary single photon.

\section{\label{sec4} Discussion and Conclusion}

\begin{figure}
	\centering
	\includegraphics[width=12cm]{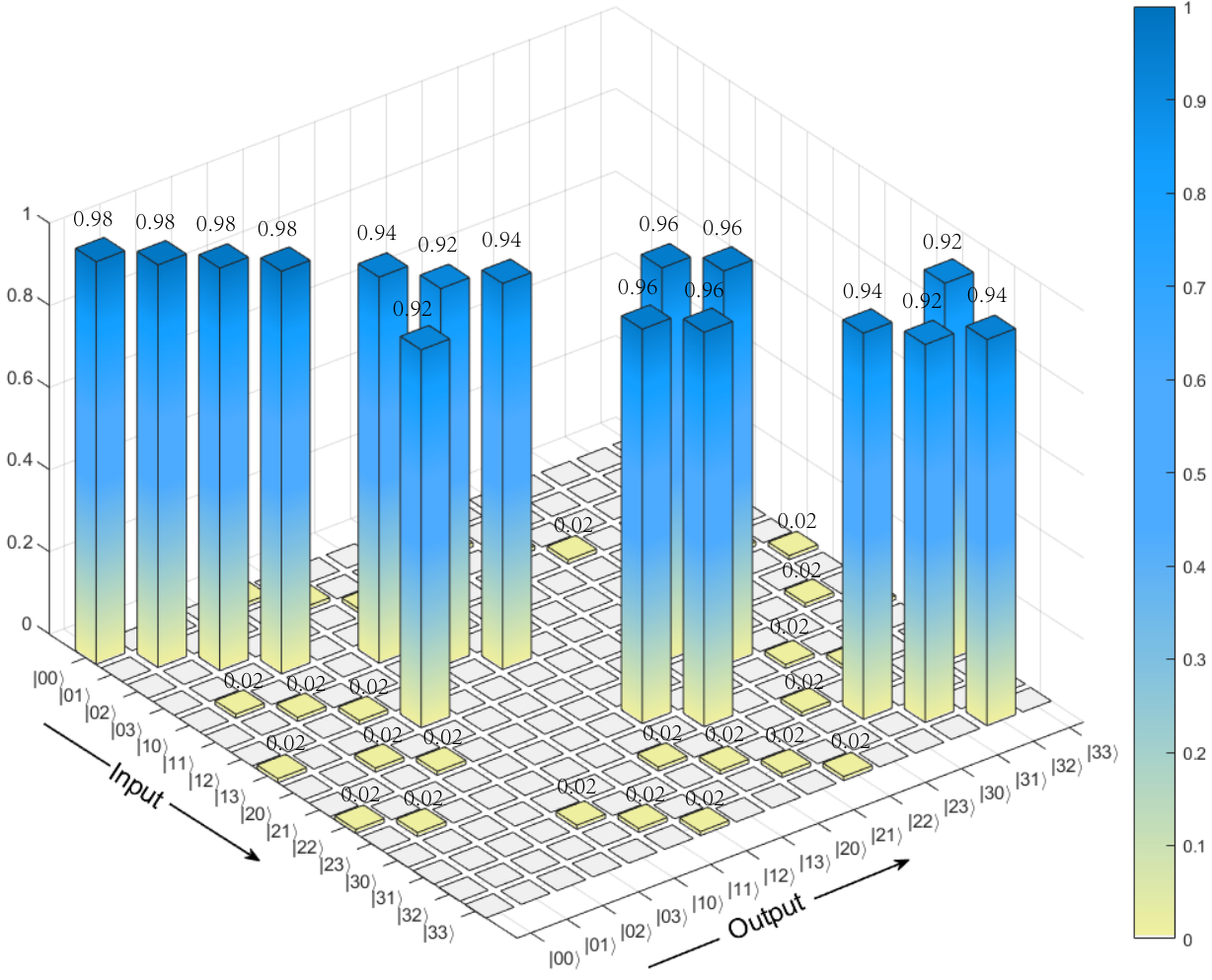}
	\caption{The probability of all computational basis states occurring after applying the 2-qudit CNOT gate to each of the computational basis states for four $\rm SiV^{-}$ centers vs the $4\times4$-dimensional
		set of input states with the condition $r_{\downarrow}=-r_{\uparrow} =0.98$ in Ref. \cite{PRA022428}.}   
	\label{fig3}   
\end{figure}

So far, we have set up definitely a 2-qudit $4\times4$-dimensional CNOT gate
for the four-SiV$^{-}$ system by virtue of
perfect optical transitions and efficacious interaction
between the auxiliary photon and four SiV$^{-}$ spins,
where the auxiliary $H$-polarization
photon carrying an input frequency $\omega$ with lower decoherence and shorter operation time
interacts each SiV$^{-}$ center in two ground states
$|\downarrow\rangle$  and $|\uparrow\rangle$
and derive various phases
$0 $ and $\pi $ from the reflection coefficients
$r_{\downarrow}(\omega)=+1$ and $r_{\uparrow}(\omega)=-1$, respectively.
Suppose that the detection efficiency of the detector $\rm D_{H1}$ (or $\rm D_{V1}$, $\rm D_{H2}$, $\rm D_{V2}$) is unity,
the efficiency of our protocol is mainly impacted from
the interacting loss, when the linear optical elements, i.e., PBSs and BSs  are ideal ignoring the errors.
The fidelity $F$ defined in the paper \cite{PhysRevLettGates} of our 2-qudit CNOT gate shown in Figure \ref{fig3}. (a) approach unity in the $4\times4$-dimensional
set of input states $\{|00\rangle,|01\rangle,|02\rangle,|03\rangle,|10\rangle,|11\rangle,|12\rangle,|13\rangle,
|20\rangle,|21\rangle,|22\rangle,|23\rangle,|30\rangle,|31\rangle,|32\rangle, $ and $|33\rangle\}$.

%
%

As external
fields and local strain are variable,
dissimilar detunings of optical transitions aiming at four SiV$^{-}$ centers,
it is essential to compensate  energetically
these detunings by adjusting external fields and
and the strain \cite{RN203,PRB205444,PRX031022}.
Furthermore, imperfect factors, i.e., a limited detuning $\Delta_{\uparrow}$ and a finite cooperativity $C$
owing to a tiny coupling $g$ coupling a spin to an optical nanocavity,
lead to a nonideal interaction process with the coefficient
$r_{d}(\omega)$ designated in Eq.(\ref{eq2}).
Especially, both the reflection coefficients $r_{\downarrow}$ and
$ -r_{\uparrow} $ with respect of various ground states equal to 0.98,
holding opposite phases and same amplitude in a condition with $C =100$,
$\Delta_{\uparrow} = 100$, and $\Delta_{c}=1$ in Ref. \cite{PRA022428}.
The Figure \ref{fig3} shows the probabilities of all computational basis states after applying the four-dimensional CNOT gate to each of the computational basis states, the minimum effective conversion rate is $P = 92.27\%$.

\begin{figure}   
	\begin{minipage}[t]{0.5\linewidth} 
		\centering   
		\includegraphics[width=7cm]{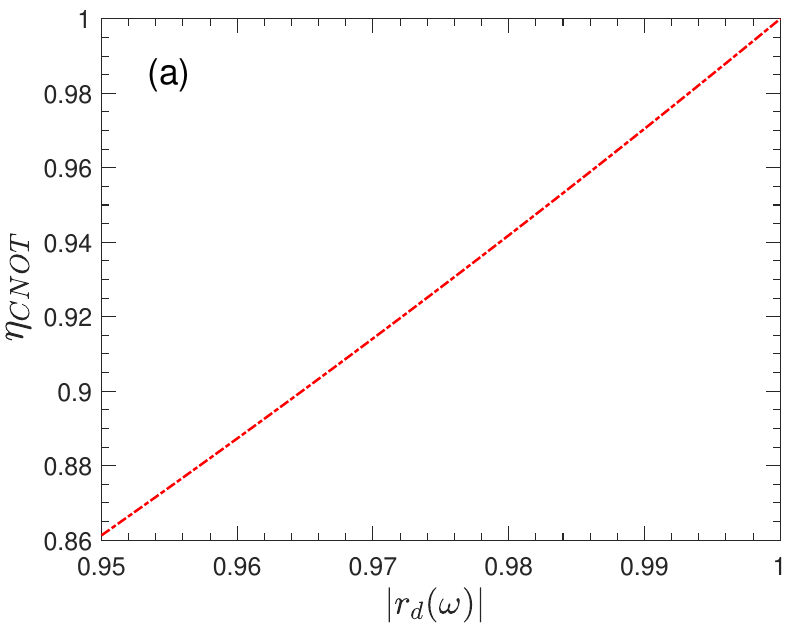}   
	\end{minipage}%
	\begin{minipage}[t]{0.5\linewidth}   
		\centering   
		\includegraphics[width=7cm]{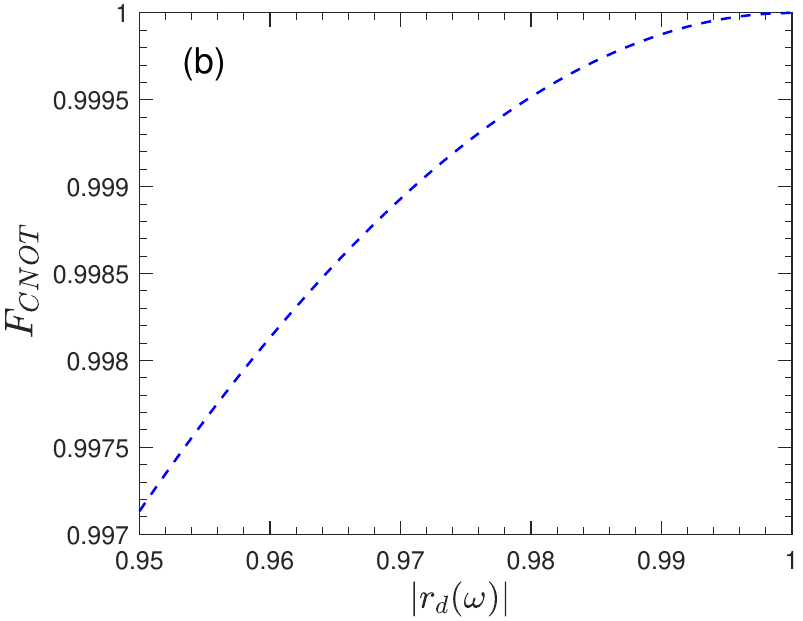}   	 
	\end{minipage}
	\caption{	(a) The efficiency and (b) fidelity versus the reflection coefficient $|r_{d}(\omega)|$. }  	\label{fig4}   
\end{figure}

The performance of the 2-qudit $4\times4$-dimensional CNOT gate is described by efficiency and fidelity, which are affected by the reflection coefficient.  Here, we treat the SiV centers as identical ones for convenience.  The reflection coefficient is shown in Equation (\ref{eq2}), an increase in the effective detunings of cavity-mode frequencies $\Delta_{\uparrow}$ will raises the reflection coefficient $|r_{d}(\omega)|$ closer to 1, as does decreasing
the dipole-transition  $\Delta_{\downarrow}$ and cavity-mode frequencies $\Delta_{c}$.                      
The efficiency of these gates can approach unity when $\Delta_{c}=\Delta_{\downarrow}=0$ and $\Delta_{\uparrow}\gg C \gg 1$.              
With parameters $(g, \gamma, \kappa) = 2\pi \times (8.4, 0.1, 28.2)$ GHz in Ref. \cite{RN200}, the cooperativity C can approximately reach 100.  Besides, the large optical
splitting $\Delta_{c} - \Delta_{\uparrow} = -99$ of two effective optical transitions
$|\downarrow\rangle\rightarrow|\downarrow'\rangle$ and
$|\uparrow\rangle\rightarrow|\uparrow'\rangle$
can be achieved in case of moderate
strain $10^{-7}$  and the magnetic field $B \sim 0.5 T$,
where the difference of Land\'{e} $g$
factors between the ground and excited states
is $\delta g = 0.06$ and meanwhile
the orbital splitting of two ground states is $\Delta \sim$ 140 GHz \cite{PRB165428}.  
And with the reflection coefficient $r_{\downarrow}=-r_{\uparrow} $ = [0.95, 0.96, 0.98, 1], the efficiencies $\eta_{C}$ = [0.8613, 0.8876, 0.9427, 1] shown in Figure \ref{fig4}. (a) and the fidelities $F_{C}$ = [0.9971, 0.9981, 0.9995, 1] shown in Figure \ref{fig4}. (b) for CNOT gate can be achieved.

Generally speaking, the decoherence of the SiV$^{-}$ center also reduces the
fidelity of the four-dimensional CNOT gate and its influence
on the fidelity coming from two aspects,
where one is the defective scattering of the auxiliary photon with
each SiV$^{-}$ spin, the other one is
the propagation of the auxiliary photon over quantum
channels between neighboring two SiV$^{-}$ spins. For the former,
the spin decoherence falls off the fidelity
defined as
$[exp(-t_{T} / T^{e}_{2})+1]/2$,
when each SiV$^{-}$
spin experiences respective decoherence \cite{PRL180403,PRA032322}.
Here $t_{T}$ ($\sim$ microsecond) is the total interaction time in each
SiV$^{-}$-cavity system and the $T^{ e}_{2}$ $(> 10 ms)$ is the electron-spin coherence time
of the SiV$^{-}$  at
100 mK \cite{PRL223602}. Therefore, the decoherence of the SiV$^{-}$
spin decreases the fidelity of the $4\times4$-dimensional CNOT gate
less than 0.005 during interaction processes in each
SiV$^{-}$. Besides,
the imperfect mode
matching between polarized state $|H\rangle$ of the auxiliary photon and the optical
transition of the SiV$^{-}$ result in an error,
decreasing the fidelity $10^{-4}$ of our CNOT gate for an available mode-matching
efficiency of 0.99 \cite{PRX021071}.

In summary, we have shown that it is possible to implement
a computation-enhanced 2-qudit  CNOT gate
for four SiV$^{-}$ spins with a auxiliary photon,
in which the control qudits encode the information on the
spin states of SiV$^{-}_{1}$ and SiV$^{-}_{2}$ centers,
and the other ones of SiV$^{-}_{3}$ and SiV$^{-}_{4}$ centers as the target qudits.
The ancillary photon in state $|H\rangle$
is used to interact with four stationary SiV$^{-}$ spins
embedded in four independent single-sided optical nanocavities and link them together.
Moreover, the CNOT gate works in a deterministic way
with perfect optical transitions and efficacious interaction
by performing the relational feed-forward operations corresponding to
the different outcomes of the single-photon detectors to be aimed at
the ancillary photon. Further, it can be potentially generalized
to other solid-state quantum system, namely, quantum dots,  nuclear magnetic resonance, atoms, and
nitrogen-vacancy
centers.
The fidelity $F$ of the 2-qudit CNOT gate for four $\rm SiV^{-}$ centers have satisfactory performances
by  exploiting existing technological conditions.
In contrast to pre-existing $2\times2$-dimensional CNOT gates CNOT gates \cite{WangTJ3,weiNV1,weiNV2,spin1,spin2,weihrspin3},
the  2-qudit $4\times4$-dimensional CNOT gate operating on solid-state quantum systems
provide not only the scalability to realize more complex quantum operations given in Eq. (\ref{eq10}),
but also the experimental flexibility to enhance practically the speed of quantum computing.

\begin{acknowledgments}
	This work was supported in part by the Natural Science Foundation of China under Contract 61901420;	
	in part by Fundamental Research Program of Shanxi Province under Contract 20230302121116.
\end{acknowledgments}

\section*{Disclosures}
The authors declare no conflicts of interest.

\section*{Data Availability Statement}
Data underlying the results presented in this paper are not publicly available at this time but may be obtained from the authors upon reasonable request.

\bibliography{apstemplate}

\end{document}